# On simultaneous fitting of nonlinear and linear rheology data: Preventing a false sense of certainty

Piyush K. Singh [*], Randy H. Ewoldt [†]

Department of Mechanical Science and Engineering, University of Illinois Urbana-Champaign, Urbana, IL 61801, USA

## Abstract

Uncertainty propagates through calculations, down to molecular scales to infer microstructural features, and up to macroscopic scales with predictive flow simulations. Here we study uncertainty quantification for sequential (two-step) versus simultaneous (all at once) fitting methods with linear and weakly-nonlinear rheological data. Using an example of a combined dataset on small-amplitude oscillatory shear (SAOS) and medium-amplitude oscillatory shear (MAOS) for a linear entangled polymer melt (cis-1,4-polyisoprene), we demonstrate with a multi-mode Giesekus model how the fit parameter uncertainties are significantly under-estimated with the sequential fit because of the neglect of model parameter correlations. These results are surprising because *weakly*-nonlinear data is only an asymptotic step away from the linear data, yet it has significant impact on calibrating the linear model parameters. Similarly, the nonlinear parameter estimates and uncertainties are impacted by considering the linear data in a simultaneous fit. To compare multi-mode spectra of nonlinear parameters (mobility parameters $\alpha_i$ from the Giesekus model), we derive new average measures based on moments of the spectra related to the high-frequency MAOS limit. The spectral averages are also sensitive to sequential versus simultaneous fitting. Our results reveal the importance of using simultaneous fitting for honest uncertainty quantification, even with weakly-nonlinear data.

[*] Current affiliation: Department of Chemical and Biomolecular Engineering, University of Illinois Urbana-Champaign, Urbana, IL 61801, USA

[†] Corresponding author: ewoldt@illinois.edu

# 1. Introduction

Fitting of rheology data using constitutive models is a common exercise for many researchers. Physically meaningful fit parameters can provide valuable material-information, e.g. the relaxation spectrum of the material [1-7], molecular weight distribution for polymer melts [8-14], and many other length and time scale related parameters for various materials [15-27]. These examples rely on model interpretability more than model accuracy, a well-known trade-off with predictive modeling and machine learning methods. Of course, accuracy may instead be preferred for the sake of prediction at the expense of interpretation [28,29], e.g. fitting rheological models for the sake of fluid mechanics flow predictions. With either objective, we desire trustworthy estimates and uncertainty quantification (UQ) of the model parameters.

Coupled datasets, for example, linear and nonlinear rheology datasets, are common scenarios, and these are subject to the option of simultaneous versus sequential fitting of model parameters. The choice arises because a number of these models [30,31], including the Giesekus model used here, are typically an extension of a linear constitutive relation via an additional nonlinear term which introduces nonlinear model parameters into the framework. As a result, although the linear rheology data model predictions involve only the linear parameters, the nonlinear data model predictions depend upon both linear and nonlinear model parameters. The linear parameters can be estimated first by fitting the linear data, followed by fitting the nonlinear data to estimate the nonlinear parameters (while keeping the linear parameters fixed at their estimates from the first step): *The sequential fit*. Alternatively, linear and nonlinear data can be fit simultaneously to estimate linear and nonlinear parameters altogether: *The simultaneous fit*.

While the sequential fit seems convenient and faster as it avoids optimizing a large parameter space simultaneously, it neglects parameter correlations, which are needed for true parameter uncertainties. This is immediately clear from the toy example of Figure 1, where the model parameters of a single mode Maxwell model are fit to the SAOS data of cis-1,4 polyisoprene melt (The data acquisition and model fitting details are described in later sections). When both model parameters are fit simultaneously, the confidence interval and associated (full) parameter uncertainties reflect the correlative effects, i.e. larger variations in both parameters are allowed. In contrast, if one parameter is kept fixed, and the other is fit to the data, we get (partial) smaller uncertainties on the parameter for model predictions with the same confidence level. Even though we get smaller model parameter uncertainties from the sequential fit, this is just a false sense of more certainty, as the parameter correlations are ignored for this case.

In this work we study how important is the distinction of sequential versus simultaneous fitting for a rheology dataset that includes linear + *weakly nonlinear* data. One might have the expectation that since we are only looking at small deviations from linearity, the linear model parameters might still dominate the fitting. If that is the case, the distinction between sequential and simultaneous fitting might be negligible for model parameter estimates and uncertainties, and one can then proceed with the more convenient sequential fitting. We show here that this is *not* the case, and model parameter uncertainties are highly impacted even when fitting weakly nonlinear rheology data. So, whenever possible, it is best to fit all linear and nonlinear model parameters simultaneously. The discussion in this paper focuses on fitting SAOS+ MAOS data specifically, but the general idea that simultaneous fitting of model parameters is the conceptually correct and more robust approach is applicable to other cases such as SAOS + steady shear, SAOS + first normal stress differences, linear shear + extension, steady nonlinear shear + thixotropic time dependence, etc.

Note that there are other subjective choices for rheology data fitting, in addition to sequential versus simultaneous fitting, which are outside the scope of this work but have been covered elsewhere. For example, in a previous study [32], we have shown that weighting choices for residuals in a least-squares fitting (e.g. no weighting, experimental data, measured or estimated data uncertainty etc.) and the data representation choices (e.g. SAOS data can be represented as dynamic moduli, compliances, complex modulus and tangent of phase difference etc.) significantly affect the parameter estimates and their uncertainty. We further showed that the conceptually correct approach is to weight the residuals by the data uncertainty, and this also mitigates the subjective effect of data representation [31,32]. Another commonly overlooked aspect is to choose a model that adequately captures the physics of the system. Accounting for missing physics in the model is non-trivial and requires assumptions [33,34], so the onus remains with the researcher to select an adequate model for the sake of interpretation.

In this work, we use experimental data weighting for least squares residuals, since this is the common approach used in the literature. We discuss how the calculations are affected when data uncertainties (from repeat measurements as an example) are available to be used for weighting instead. For the model, we use the Giesekus model which captures the trend and signs of the SAOS+MAOS material functions, and hence, adequately captures the physics of the system.

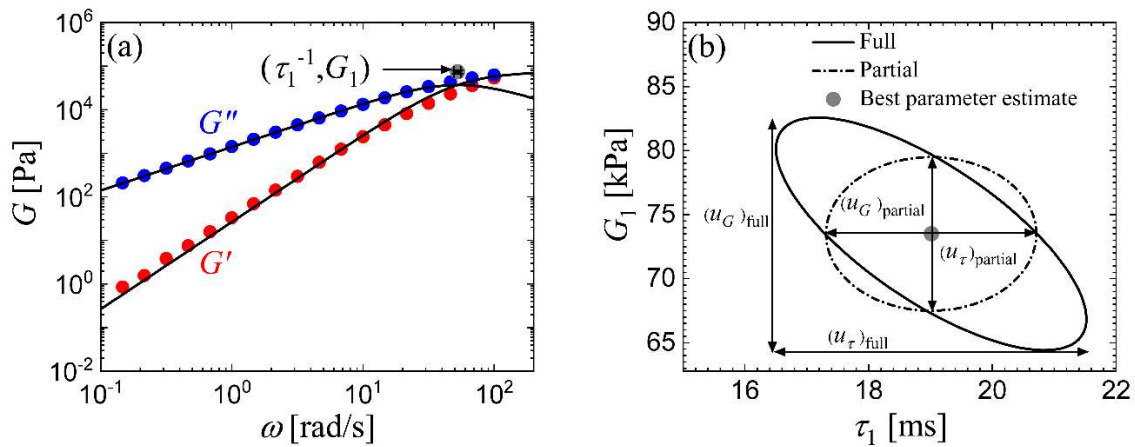

**FIG. 1.** (a) A single-mode Maxwell model fit to SAOS data of a cis-1,4-polyisoprene melt (using data weighted least-squares fitting of Eq. (24)) along with the fit parameter estimates, and (b) the fit parameter estimates along with the 68.3% confidence ellipses. The full uncertainty in parameters accounts for all correlations. Keeping one parameter fixed ignores the correlations and provides only a partial uncertainty, which although smaller than the full uncertainty, is wrong nonetheless for microstructural inference purposes.

## 2. Background

### 2.1 Model parameter uncertainties in fitting

Fitting of data is typically performed using the least-squares method which in its most generalized form minimizes the residual sum of squares (RSS) [35-38] which has the form

$$\text{RSS} = \sum_{i=1}^{N} \left( \frac{S_i(x_i) - \hat{S}_i(x_i, \vec{p})}{w_i} \right)^2 \qquad (1)$$

where $S$ are the measured signals, $\hat{S}$ are model predictions, $w$ are the residual weights, $x$ is the independent variable (for example, angular frequency $\omega$ for oscillatory shear measurements), $\vec{p}$ is the set of model parameters, and $N$ is the number of measurement points. It is common in the rheology literature to assume an error model for $w$, for example, the two commonly used options are, $w = 1$ (constant error model) and $w = S$ i.e. the experimental data at each point (constant relative error model) [39,40]. However, weighting residuals by the true data uncertainty, i.e. $w = u_S$ where $u_S$ is the experimental uncertainty in data $S$, is the most fundamental choice [35-37,41]. In fact, weighting by the data uncertainty also reduces the effect of data representation on fitting results, as we showed in our previous work [32].

When RSS is defined using Eq. (1) with data uncertainty weighting, i.e. $w = u_S$, the corresponding uncertainty in model parameter $p_i$ for an $R$-dimensional model parameter space $\vec{p} = \{p_1, p_2, ..., p_i, ...., p_R\}$, is given as [41]

$$\left( u_{p_i} \right)_{\text{full}} = \left\{ \sqrt{2 \left[ (\nabla\nabla \text{RSS})^{-1} \right]_{ii}} \right\}_{\{\vec{p}_0\}}. \qquad (2)$$

Here $\{\vec{p}_0\}$ is the optimum obtained by minimizing RSS and $\nabla\nabla\text{RSS}$ is a second order tensor represented in matrix form as

$$\nabla\nabla \text{RSS} = \frac{\partial^2 \text{RSS}}{\partial p_i \partial p_j} = \begin{pmatrix} \dfrac{\partial^2 \text{RSS}}{\partial p_1^2} & \cdots & \dfrac{\partial^2 \text{RSS}}{\partial p_1 \partial p_R} \\ \vdots & \ddots & \vdots \\ \dfrac{\partial^2 \text{RSS}}{\partial p_1 \partial p_R} & \cdots & \dfrac{\partial^2 \text{RSS}}{\partial p_R^2} \end{pmatrix}. \tag{3}$$

(Note that in Eq. (2), $\left[(\nabla\nabla \text{RSS})^{-1}\right]_{ii}$ represents a single diagonal element of the matrix, and it is not to be confused with the summation under Einstein notation). If all parameter correlations are ignored in Eq. (2), i.e. $\dfrac{\partial^2 \text{RSS}}{\partial p_i \partial p_j} = 0$ ($i \neq j$), then we get only the partial uncertainty

$$\left(u_{p_i}\right)_{\text{partial}} = \left\{ \sqrt{2\left(\dfrac{\partial^2 \text{RSS}}{\partial p_i^2}\right)^{-1}} \right\}_{\{\bar{p}_0\}}. \tag{4}$$

The partial uncertainty of Eq. (4) (based on the inverse of the diagonal element) is smaller than the uncertainty of Eq. (2) (based on the diagonal element of the inverse of a matrix), as it ignores all parameter correlations. We use the subscript 'partial' to represent that it is not correct to use this for the full and true uncertainty of parameters. In sequential fitting of rheology datasets, some correlations are ignored while others are accounted for. So, the actual uncertainties of sequential fitting will lie somewhere between the two limits of Eqs. (2) and (4). Also, note that the best fit parameter estimates $\{\bar{p}_0\}$ for sequential and simultaneous fitting will be different.

Fit parameter uncertainty calculations of Eqs. (2) and (4) require knowledge of experimental data uncertainties. In this work, however, we use a single dataset and assumed $w = S$ (constant relative error model), for which the data uncertainties are known only up to a multiplicative factor i.e. $u_S = c \times w$. Here, the multiplicative factor $c$ is estimated as [32,35]

$$c = \sqrt{\text{RSS}_{\min}/(N-R)} \tag{5}$$

where $N$ is the total number of data points, and $R$ is the number of model parameters used for fitting.

## 2.2 Medium-amplitude oscillatory shear (MAOS)

For the purpose of demonstration, we use a MAOS dataset, which generates in addition to the linear viscoelastic moduli $G'(\omega)$ and $G''(\omega)$, four asymptotically (i.e. weakly) nonlinear material functions: $[e_1](\omega)$, $[v_1](\omega)$, $[e_3](\omega)$ and $[v_3](\omega)$ in terms of the Chebyshev framework of Ewoldt and Bharadwaj [42].

Why do we choose MAOS instead of large-amplitude oscillatory shear (LAOS)? LAOS is more nonlinear and therefore contains more information, is closer to industrial processing conditions [43], and an active area of research [43-54]. However, it requires careful experimentation to avoid artifacts and instabilities [55-62], may involve more complex mathematical frameworks [43,48,51,63-69] to define physically meaningful material functions [43,51,68,70-73], and typically no tractable analytical solution is available for calibrating model parameters to LAOS. To mitigate these challenges of LAOS, it is sometimes preferred to look at nonlinear oscillatory shear only in the limit of asymptotic deviation from linearity [42], i.e. the MAOS regime. Using a power-series expansion of stress harmonics in terms of total strain, we get four nonlinear data signals in addition to the two linear signals ($G'$ and $G''$), with a physical interpretation available for each, and where a number of analytical model predictions have been made that can be used for fitting [16,30,31,71,74]. Experimentally, MAOS is defined within a sweet spot of strain amplitudes; the amplitude is large enough to resolve nonlinearities above the noise in the signals, yet small enough that the power-law expansion coefficients adequately describe the nonlinearities and higher-order effects can be neglected [30,42,75-79].

The weakly-nonlinear regime of MAOS is particularly interesting for our study of parameter uncertainty quantification. Being only an asymptotic step away from linearity, we have a dataset where the linear model parameters might be expected to dominate the fitting, potentially reducing the effect of simultaneous versus sequential fitting (although we show that this choice has a pronounced effect even with a MAOS dataset).

In this work we considered the more common strain controlled protocol of MAOS, i.e. MAOStrain [42]. Any reference to MAOS in the rest of the work will imply MAOStrain (rather than stress-control). For MAOS, the shear strain input is represented as [70]

$$\gamma(t) = \gamma_0 \sin(\omega t) \tag{6}$$

where $\gamma_0$ is the strain amplitude and $\omega$ is the angular frequency. This in turn imposes an orthogonal strain rate given by

$$\dot{\gamma}(t) = \gamma_0 \omega \cos(\omega t) \tag{7}$$

where $\gamma_0 \omega$ is the strain-rate amplitude. Using the Chebyshev representation of Ewoldt and Bharadwaj [42], the power series expansion of the time domain shear stress in the MAOS regime is given as

$$\begin{aligned}\sigma(t;\gamma_0,\omega) = &\gamma_0 \{G'(\omega)\sin\omega t + G''(\omega)\cos\omega t\} \\ &+\gamma_0^3 \begin{Bmatrix} [e_1](\omega)\sin\omega t + \omega[v_1](\omega)\cos\omega t \\ -[e_3](\omega)\sin 3\omega t + \omega[v_3](\omega)\cos 3\omega t \end{Bmatrix} \\ &+O(\gamma_0^5). \end{aligned} \tag{8}$$

In addition to the two familiar linear viscoelastic material functions $G'(\omega)$, $G''(\omega)$, we have four nonlinear material functions $[e_1](\omega)$, $[v_1](\omega)$, $[e_3](\omega)$, $[v_3](\omega)$, where "*e*" represents elastic and "*v*" represents viscous nonlinearity, and the subscript represents the integer harmonic of the input frequency at which the nonlinearity occurs. While the linear viscoelastic moduli are always positive, the four nonlinear measures can be either positive or negative, with material property interpretations that can be assigned to a combination of signs [42]. Note that some previous studies use lumped measures rather than the four separate MAOS measures [68,69,77]. While the lumped measures are useful, they convolute the effect of the four signals and take away the sign information of the four measures. Signal decomposition with signs has been shown to be an extremely valuable tool for microstructural inference and model development [16,30,31,80].

Experimental measurements involve strain-amplitude sweeps at each angular frequency of interest, and from which the first and third harmonic elastic and viscous stresses are calculated by taking the Fourier transform of the stress waveform under steady-alternance conditions. In the MAOS regime, the data is described using the following relations:

$$\sigma_1'(\gamma_0,\omega) = \gamma_0 G'(\omega) + \gamma_0^3 [e_1](\omega) + O(\gamma_0^5), \tag{9}$$

$$\sigma_1''(\gamma_0,\omega) = \gamma_0 G''(\omega) + \omega\gamma_0^3 [v_1](\omega) + O(\gamma_0^5), \tag{10}$$

$$\sigma_3'(\gamma_0,\omega) = -\gamma_0^3 [e_3](\omega) + O(\gamma_0^5), \tag{11}$$

$$\sigma_3''(\gamma_0,\omega) = \omega\gamma_0^3 [v_3](\omega) + O(\gamma_0^5). \tag{12}$$

The material functions $G'(\omega)$, $G''(\omega)$, $[e_1](\omega)$, $[v_1](\omega)$, $[e_3](\omega)$, $[v_3](\omega)$ are obtained by fitting the relationships of Eqs. (9)-(12) within the experimental sweet spot of MAOS, below which the data is too noisy, and above which the higher order terms are too large. We refer the reader to our previous works to look at examples for the aforementioned procedure [42,79,81]. As an alternative to the tedious process of performing strain-sweeps at each angular frequency for generating MAOS data, we proposed a frequency-sweep MAOS method where the MAOS data can be generated by performing frequency-sweeps at two different strain amplitudes [79]. However, for this work, the data was generated using the traditional method of strain-sweeps. Note that while Eqs. (9)-(12) rely on integer power series expansions, typical of a vast number of material classes, non-integer scaling has also been observed for certain materials [82-84]. The cis-1,4-polyisoprene melt used in this work was confirmed to show the integer scaling of Eqs. (9)-(12) in a previous study [81].

**2.3 Multimode Giesekus model**

Our system of interest is an entangled linear polymer melt (details in Section 3.1) where the primary stress relaxation occurs via reptation [85-91]; although a more accurate description requires the consideration of other secondary relaxation processes such as contour-length fluctuations, constraint release, and high frequency Rouse modes [85,92-94]. Several sophisticated theoretical frameworks (for a review, see for example [85,92]) have been developed for combining these processes, and they have been previously applied to polyisoprene melt data (linear regime) as well [95,96]. We showed in a previous work [32] that the terminal regime linear viscoelasticity of the cis-1,4-polyisoprene melt used in this work can be adequately described by Doi's modified reptation model to include contour-length fluctuations [85,94,97-101]. However, closed form analytical solutions of MOAS material functions are only available for the original Doi-Edwards reptation model [31,91,102], which can be considered as the most basic MAOS model grounded in microstructural physics for entangled polymer melts. Additionally, this model does not have a nonlinear fitting parameter for MAOS data prediction, and therefore is not useful for demonstration purposes of this work.

For demonstrating the ideas of sequential versus simultaneous fitting of MAOS data, we instead use a multi-mode Giesekus model. We avoid the deeper questions about model selection [33], which quantitatively assess credibility of different models for a given dataset, and directly use the Giesekus model. The Giesekus model has an analytical solution for MAOS and provides the simplicity of having a single nonlinear parameter for each mode, while having a sufficiently complex mathematical structure where the nonlinear parameter appears both as a front factor and within frequency-dependent terms. Additionally, the Giesekus model adequately captures the trend and sign changes of the observed MAOS material functions as shown later.

The Giesekus model was originally developed [103,104] to model the nonlinear rheological behavior of polymers in solution, however, it has since been used for other systems including wormlike micelles [105] and polymer melts [106]. The single mode Giesekus model is represented as a tensorial equation in polymeric stress $\underline{\underline{\sigma}}$ as

$$\underline{\underline{\sigma}} + \tau \underline{\underline{\sigma}}_{(1)} + \frac{\alpha}{G} \underline{\underline{\sigma}} \cdot \underline{\underline{\sigma}} = G\tau \underline{\underline{\dot{\gamma}}} \tag{13}$$

where $\underline{\underline{\sigma}}_{(1)}$ is the upper convected derivative of stress defined as

$$\underline{\underline{\sigma}}_{(1)} = \frac{\partial}{\partial t}\underline{\underline{\sigma}} + (\underline{v} \cdot \underline{\nabla})\underline{\underline{\sigma}} - (\underline{\nabla}\underline{v})^T \cdot \underline{\underline{\sigma}} - \underline{\underline{\sigma}} \cdot (\underline{\nabla}\underline{v}) \tag{14}$$

where the velocity gradient tensor $\underline{\nabla}\underline{v}$ has components $\nabla_i v_j = \partial v_j / \partial x_i$. The parameters $\tau$ and $G$ are the relaxation time and the relaxation modulus which are related to the zero-shear viscosity as $\eta_0 = G\tau$.

The nonlinear dimensionless parameter $\alpha$ is known as the mobility factor (or the drag anisotropy coupling parameter) and it relates the deformation of the polymer to the drag force. Its value varies from $0 \leq \alpha \leq 1$, where $\alpha = 0$ represents the case when the drag force is independent of the configuration of the polymer and for this case, Eq. (13) reduces to the upper-convected Maxwell model. The case of $\alpha =1$ represents maximum coupling between the configuration of the polymer and the drag force. Such a case represents a drag force which is anisotropic and highly variable with respect to the configuration of the polymer. The mobility factor has additional significance for wormlike micellar solutions where it distinguishes between different nonlinear effects. For that particular system, a value of up to 0.5 is associated with shear thinning while greater values have been associated with shear banding as shown in the work of Helgeson, Reichert, Hu and Wagner [105].

Eq. (13) represents the polymeric stress contribution due to a single mode of deformation characterized by the three parameters: $\tau$, $G$ and $\alpha$. The Giesekus model can be extended to a multimode version [107-109]. For such an extension, each mode has its own set of the three parameters with the stress contribution governed by

$$\underline{\underline{\sigma}}_i + \tau_i \underline{\underline{\sigma}}_{i(1)} + \frac{\alpha_i}{G_i} \underline{\underline{\sigma}}_i \cdot \underline{\underline{\sigma}}_i = G_i \tau_i \underline{\underline{\dot{\gamma}}}. \qquad (15)$$

In case of a $K$ mode representation, each mode will have a separate governing equation for stress as shown above and the total stress is given by:

$$\underline{\underline{\sigma}} = \sum_{i=1}^{K} \underline{\underline{\sigma}}_i. \qquad (16)$$

For a sinusoidal deformation input represented by Eq. (6), the MAOS material functions for a single mode Giesekus model were derived by Gurnon and Wagner [110], which were then translated into the Chebyshev framework of Ewoldt and Bharadwaj [42] (as used herein) by Bharadwaj and Ewoldt [30] (their Eqs. 34-37). These single-mode predictions can then be generalized to a multimode Giesekus model framework (assuming stress from various modes are additive) as

$$G'(\omega) = \sum_{i=1}^{K} G_i \frac{(\tau_i \omega)^2}{1+(\tau_i \omega)^2} \tag{17}$$

$$G''(\omega) = \sum_{i=1}^{K} G_i \frac{\tau_i \omega}{1+(\tau_i \omega)^2} \tag{18}$$

$$[e_1](\omega) = \sum_{i=1}^{K} G_i \alpha_i \frac{(\tau_i \omega)^4 \left(-21 - 41(\tau_i \omega)^2 - 8(\tau_i \omega)^4 + 4\alpha_i \left(4 + 7(\tau_i \omega)^2\right)\right)}{4\left(1+(\tau_i \omega)^2\right)^3 \left(1+4(\tau_i \omega)^2\right)} \tag{19}$$

$$[v_1](\omega) = \sum_{i=1}^{K} -G_i \tau_i \alpha_i \frac{(\tau_i \omega)^2 \left(9 + 11(\tau_i \omega)^2 - 10(\tau_i \omega)^4 + 2\alpha_i \left(-3 - (\tau_i \omega)^2 + 8(\tau_i \omega)^4\right)\right)}{4\left(1+(\tau_i \omega)^2\right)^3 \left(1+4(\tau_i \omega)^2\right)} \tag{20}$$

$$[e_3](\omega) = \sum_{i=1}^{K} -G_i \alpha_i \frac{(\tau_i \omega)^4 \left(-21 - 30(\tau_i \omega)^2 + 51(\tau_i \omega)^4 + 4\alpha_i \left(4 - 17(\tau_i \omega)^2 + 3(\tau_i \omega)^4\right)\right)}{4\left(1+(\tau_i \omega)^2\right)^3 \left(1+4(\tau_i \omega)^2\right)\left(1+9(\tau_i \omega)^2\right)} \tag{21}$$

$$[v_3](\omega) = \sum_{i=1}^{K} G_i \tau_i \alpha_i \frac{(\tau_i \omega)^2 \left(-3 + 48(\tau_i \omega)^2 + 33(\tau_i \omega)^4 - 18(\tau_i \omega)^6 + \alpha_i \left(2 - 48(\tau_i \omega)^2 + 46(\tau_i \omega)^4\right)\right)}{4\left(1+(\tau_i \omega)^2\right)^3 \left(1+4(\tau_i \omega)^2\right)\left(1+9(\tau_i \omega)^2\right)}$$

(22)

The linear viscoelastic material functions in Eqs. (17)-(18) depend only upon $\{\tau_i, G_i\}$ i.e. the relaxation spectrum of the material. In contrast, all four MAOS measures given by Eqs. (19)-(22) depend upon all three parameters of each mode: $\{\tau_i, G_i, \alpha_i\}$.

The mathematical structure of the Giesekus model MAOS predictions (and many MAOS predictions in general) then enables the two options for fitting, as discussed earlier, and shown in Fig. 2: the more commonly employed *sequential fit* [16,78,107,109-112] and the conceptually correct *simultaneous fit* [113-115] which accounts for all model parameter correlations. In what follows, we demonstrate that the two fitting schemes lead to significant differences which cannot be ignored, even when looking at a dataset with asymptotic deviations from linearity.

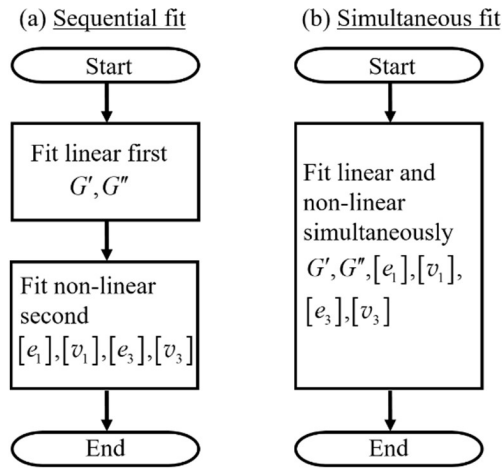

**FIG. 2.** The two ways of fitting SAOS+MAOS data: (a) sequetial fit (two-step for MAOS) where the linear SAOS data is fit first to obtain the linear parameters, which are then fixed at their best estimates during subsequent MAOS data fitting, and (b) simultaneous (all-at-once) fit, in which linear and nonlinear data are fit simultaneously to obtain linear and nonlinear parameters.

### 3. Material, dataset, and fitting methods

The SAOS+MAOS dataset used here was provided by Dr. N. Ashwin Bharadwaj, some of which is published [81], while the rest of it is reported here for the first time. The full details of the material and rheometry methods can be found in the published work of Bharadwaj and Ewoldt [81], which we summarize below, along with the fitting methods used for this work.

### 3.1 Material: Cis-1,4-polyisoprene melt

The material used for collecting the dataset was a linear and well-entangled 1,4-polyisoprene melt (high cis content) with a molecular weight, $M = 54$ K ($M_e = 6.4$ K [116], $M/M_e \approx 8.4$, hence well entangled), and a polydispersity ratio $Đ \sim 1.2$. This material is supplied by Kuraray America Corporation under the trade name LIR50 (where LIR stands for liquid isoprene rubber). This material displays an extended terminal regime accessible at room temperature, making it a preferable material for calibrating phase angles close to 90º [117]. The linear oscillatory shear [95,118], and linear and nonlinear start-up of shear [95] rheology of this material has been characterized for a range of molecular weights. Bharadwaj and Ewoldt [81] utilized the extended terminal regime of this material to validate their predictions of the MAOS terminal regime scaling and interrelations.

### 3.2 Rheometry methods and data

Rheometry measurements for this experimental dataset [81] were performed on a separated motor-transducer rheometer (TA Instruments ARES-G2) using a cone and plate geometry (50 mm diameter and 2° cone angle). A cone and plate geometry ensures spatially homogeneous simple shear deformation within the material [58]. A large diameter was chosen here for better torque sensitivity at low frequencies. The experiments were carried out at a temperature of 25ºC maintained by a Peltier system in the lower plate.

The linear viscoelastic response shown in Fig. 3(a) is the same as in the published work of Bharadwaj and Ewoldt [81] (their Fig. 3). A fixed strain amplitude of $\gamma_0 = 1\%$ was used for the SAOS measurements. The chosen strain amplitude was confirmed to lie in the linear viscoelastic regime based on the plateaus in $G'$ and $G''$ during strain-amplitude sweeps. The two linear viscoelastic moduli are positive throughout and show the expected terminal regime scaling behavior of $G' \sim \omega^2$ and $G'' \sim \omega$ at low frequencies. The instrument torque resolution limit $T_{\min}$ sets the minimum measurable modulus as [42,55]

$$G_{\min} = \frac{F_\sigma T_{\min}}{\gamma_0}, \tag{23}$$

where $F_\sigma = 3/(2\pi R^3)$ is the stress conversion factor for the cone-plate geometry with radius $R$ [58], resulting in the low-torque limit as shown by the horizontal line in Fig. 3(a). Here the manufacturer specified limit of $T_{\min} = 0.05$ μNm was used.

The MAOS response is shown in Fig. 3(b)-(e). The details of measurement of these asymptotic nonlinearities are given in the original publication by Bharadwaj and Ewoldt [81]. The low frequency data reported here is directly taken from their work but the high frequency data ($\omega \geq 1.77$ rad/s) is reported here for the first time. The inclusion of high frequency data beyond the terminal regime makes sure that an adequate model for this dataset has more than one mode; a better comparison between sequential and simultaneous fitting is made for a multi-mode model.

Note that the number of SAOS and MAOS data points are different here, as SAOS data was collected beforehand through a frequency-sweep, and then the MAOS datapoints were collected using strain-sweeps at a higher point density than SAOS. However, SAOS and MAOS material functions can be determined together from strain-sweeps using Eqs. (9)-(12), or with two frequency-sweeps for frequency-sweep MAOS. Here, we are using SAOS and MAOS material functions determined separately to mimic the case of mixing different protocols for acquiring linear and nonlinear data, such as SAOS and steady shear flow data. It is also to be noted that the model parameter estimates and their uncertainties depend upon the number of data points used. When fitting multiple datasets, a sufficient number of points for each dataset has to be used to ensure the fitting is not biased by any single dataset.

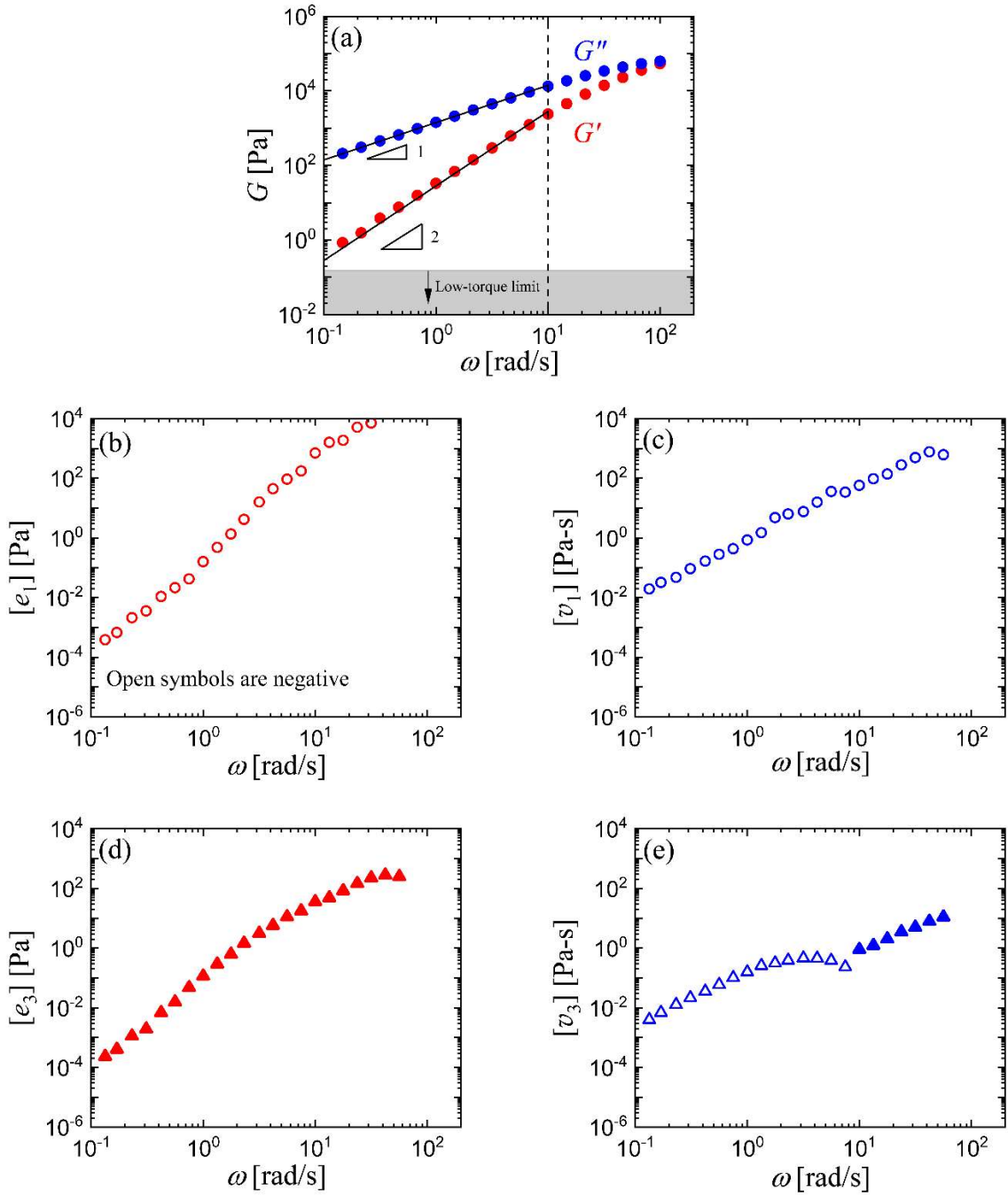

**FIG. 3.** Oscillatory shear data on a cis-1,4-polyisoprene melt: (a) SAOS, (b)-(e) MAOS material functions. The linear viscoelastic response in (a) is the same as in the published work of Bharadwaj and Ewoldt [81], and shows terminal regime scaling behavior of $G' \sim \omega^2$ and $G'' \sim \omega$ at small frequencies. The low-torque limit is calculated using Eq. (23). For this dataset, $[v_3]$ changes sign from negative at small frequencies to positive at large frequencies. The other three MAOS material functions retain their sign over the whole frequency domain. However, they have different signs: $[e_1]$ and $[v_1]$ are always negative while $[e_3]$ is always positive.

MAOS material functions can be either positive or negative, and additionally, change sign over the frequency domain, unlike linear viscoelastic moduli which are always positive. For this dataset, the only sign change is observed for $[v_3]$ which is negative at small frequencies and positive at large frequencies. The remaining three MAOS material functions retain their sign over the tested frequency domain. However, they have different signs: $[e_1]$ and $[v_1]$ are always negative while $[e_3]$ is always positive.

The signs and magnitudes of first- and third-harmonic MAOS material functions have interpretation for viscoelastic characteristics of this material [42]. More specifically, a negative $[e_1]$ implies that this material is elastic-softening, which is a typical characteristic of entangled polymer melts, implying that with the application of larger deformations, the material softens elastically. A positive $[e_3]$ informs that this elastic softening is driven by large strain-rates for this range of angular frequencies. A negative $[v_1]$ implies that the material is viscous-thinning, which means that with the application of larger deformations, viscosity of the material decreases. The sign change in $[v_3]$ informs that this viscous-thinning is driven by large strain-rates $\dot{\gamma}$ for $\omega \leq 7.5$ rad/s, and by large strains $\gamma$ for $\omega > 7.5$ rad/s. These interpretations are consistent with a nonlinear viscoelastic constitutive model that shows softening/thinning due to large strain rates at small frequencies, and large strains at large frequencies. The Giesekus model has these characteristics and is therefore suitable for fitting this dataset.

### 3.3 Fitting methods

We used a standard least-squares method with experimental data weighting for fitting. As mentioned previously in Section 2.1, whenever available, the data uncertainty is the fundamental weighting for residuals. However, since we are looking at a single measurement dataset here, we choose the next best weighting, the experimental data itself (this makes sure that the data varying over orders of magnitude is fairly accounted for in fitting [32,119]).

Sequential fitting required fitting the linear data by minimizing Eq. (24) first to obtain the linear relaxation spectrum $\{\tau_i, G_i\}$ followed by fitting nonlinear data by minimizing Eq. (25) to obtain nonlinear parameters $\{\alpha_i\}$. On the other hand, for simultaneous fitting, linear and nonlinear data were fit simultaneously by minimizing Eq. (26) to obtain linear and nonlinear parameters $\{\tau_i, G_i, \alpha_i\}$ together.

$$\text{RSS}_{\text{linear(sequential)}}(\{\tau_i, G_i\}) = \sum_{j=1}^{N_1} \left\{ \left( \frac{G'_j(\omega_j) - \hat{G}'_j(\omega_j; \{\tau_i, G_i\})}{G'_j(\omega_j)} \right)^2 + \left( \frac{G''_j(\omega_j) - \hat{G}''_j(\omega_j; \{\tau_i, G_i\})}{G''_j(\omega_j)} \right)^2 \right\}$$

(24)

$$\text{RSS}_{\text{nonlinear(sequential)}}(\{\alpha_i\}) = \sum_{j=1}^{N_2} \left\{ \begin{array}{l} \left( \dfrac{[e_1]_j(\omega_j) - [\hat{e}_1]_j(\omega_j; \{\alpha_i\})}{[e_1]_j(\omega_j)} \right)^2 + \left( \dfrac{[v_1]_j(\omega_j) - [\hat{v}_1]_j(\omega_j; \{\alpha_i\})}{[v_1]_j(\omega_j)} \right)^2 \\ + \left( \dfrac{[e_3]_j(\omega_j) - [\hat{e}_3]_j(\omega_j; \{\alpha_i\})}{[e_3]_j(\omega_j)} \right)^2 + \left( \dfrac{[v_3]_j(\omega_j) - [\hat{v}_3]_j(\omega_j; \{\alpha_i\})}{[v_3]_j(\omega_j)} \right)^2 \end{array} \right\}$$

(25)

$$\text{RSS}_{\text{global(simultaneous)}}(\{\tau_i, G_i, \alpha_i\}) =$$

$$\sum_{j=1}^{N_1} \left\{ \left( \frac{G'_j(\omega_j) - \hat{G}'_j(\omega_j; \{\tau_i, G_i\})}{G'_j(\omega_j)} \right)^2 + \left( \frac{G''_j(\omega_j) - \hat{G}''_j(\omega_j; \{\tau_i, G_i\})}{G''_j(\omega_j)} \right)^2 \right\} +$$ (26)

$$\sum_{j=1}^{N_2} \left\{ \begin{array}{l} \left( \dfrac{[e_1]_j(\omega_j) - [\hat{e}_1]_j(\omega_j; \{\tau_i, G_i, \alpha_i\})}{[e_1]_j(\omega_j)} \right)^2 + \left( \dfrac{[v_1]_j(\omega_j) - [\hat{v}_1]_j(\omega_j; \{\tau_i, G_i, \alpha_i\})}{[v_1]_j(\omega_j)} \right)^2 \\ + \left( \dfrac{[e_3]_j(\omega_j) - [\hat{e}_3]_j(\omega_j; \{\tau_i, G_i, \alpha_i\})}{[e_3]_j(\omega_j)} \right)^2 + \left( \dfrac{[v_3]_j(\omega_j) - [\hat{v}_3]_j(\omega_j; \{\tau_i, G_i, \alpha_i\})}{[v_3]_j(\omega_j)} \right)^2 \end{array} \right\}$$

The "hat" notation in Eqs. (24)-(26) represents the model predictions, and $N_1 = 18$ and $N_2 = 22$ are the number of frequencies at which SAOS and MAOS data were collected respectively. For simultaneous fitting, Eq. (26) provides the global RSS, while for sequential fitting, the global RSS is calculated as the sum of linear and nonlinear contributions from Eq. (24) and Eq. (25) respectively.

The optimization of RSS in Eqs. (24)-(26) was implemented using the Levenberg-Marquardt algorithm [35,38] within OriginPro 2017 software. The fitting was performed with a maximum number of iterations = 400 until a tolerance of $10^{-9}$ was achieved on the RSS. Here, tolerance is defined as the difference between the RSS of the last iteration and the current iteration divided by their sum. Usually the fits converged within ten iterations.

## 4. Results and discussion

### 4.1. Goodness of fit comparison

We performed sequential and simultaneous fitting for the data of Fig. 3 using up to $K = 5$ modes of the Giesekus model. The optimized global RSS (i.e. sum of minimum of Eqs. (24)-(25) for sequential fit and minimum of Eq. (26) for simultaneous fit) versus number of modes is shown in Fig. 4. Overall, optimized global RSS decreases with the number of modes for both sequential and simultaneous fits because the increase in the number of fit parameters provides additional degrees of freedom for fitting the data. In both cases, the goodness of fit seems to saturate as evident by the plateau in optimized global RSS at larger number of modes.

Simultaneous fitting provides better fits compared to sequential fitting as evident by lower optimized global RSS in Fig. 4. This is also clearly visible in terms of data fits as demonstrated for a three-mode Giesekus model in Fig. 5. The fitting is dominated by the MAOS material functions because: (1) there are four MAOS signals as compared to two SAOS signals, and (2) MAOS data is non-trivial showing various trends and sign-changes while SAOS data shows a simple Maxwell-like response. Addition of each mode provides only one more parameter for nonlinear data fitting in the sequential fitting scheme, while three more parameters are available per mode in simultaneous fitting. As a result, simultaneous fitting is more optimized for the same number of modes.

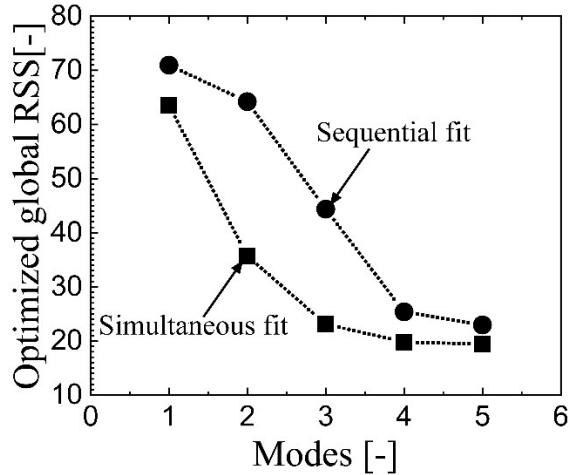

**FIG. 4.** The optimized global RSS versus number of modes for the Giesekus model fit to the data of Fig. 3, using both sequential and simultaneous fitting schemes of Fig. 2. Optimized global RSS is calculated as the sum of minimum of Eqs. (24)-(25) for sequential fit and minimum of Eq. (26) for simultaneous fit .

Based on these results, one must naturally prefer simultaneous fitting over sequential fitting for an improved goodness of fit. Also, as we have already shown, simultaneous fitting provides correct model parameter uncertainty by accounting for all parameter correlations. We now look at how significant the differences are for the two schemes in the context of the latter point. For the purpose of demonstration, we look at the three-mode Giesekus model where the goodness of fit starts to saturate, and where a significant number of model parameter correlations are ignored under sequential fitting.

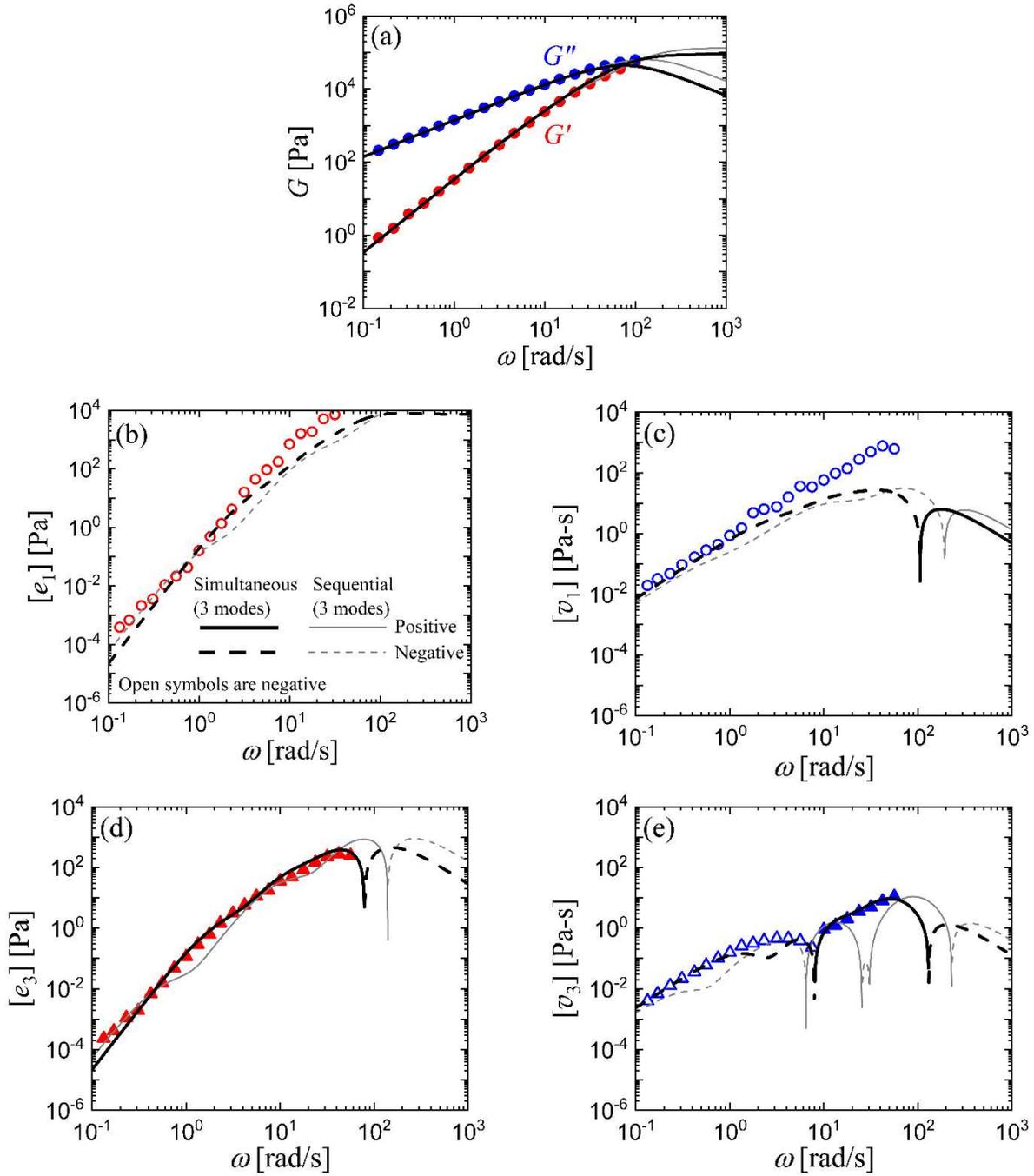

**FIG. 5.** Three-mode Giesekus model fit to the data of Fig. 3 for both sequential and simultaneous fit schemes of Fig. 2. The corresponding fit parameters along with their uncertainties are shown in Fig. 6. Model fits for other number of modes, corresponding to Fig. 4, are shown in the supplementary material.

## 4.2 Simultaneous fit gives honest and correct model parameter uncertainty

Figure 5 shows the data fit to the three-mode Giesekus model using both sequential and simultaneous fitting. The associated model parameter estimates and uncertainties are shown in Fig. 6 and listed in Table 1. While the parameter estimates for the two schemes in Fig. 6 are different because of different optimizing functions, the general trend of parameters, for example, the variation of $G_i$ with $\tau_i$, are similar.

The apparent model parameter uncertainties of sequential fitting are much smaller, as evident by a maximum relative uncertainty (= 100 × uncertainty/fit value) of 47 % in the nine parameters, while eight of the parameters have a relative uncertainty below 21.2 %. In contrast, the relative uncertainties are always larger for simultaneous fitting (except for $\tau_3$). In fact, the uncertainties in $G_i$ and $\alpha_i$ of second and third modes (the longer relaxation times) are larger than 100%, i.e. the uncertainties are larger than the fit values themselves!

The reason we get larger uncertainties, especially in $G_i$ and $\alpha_i$, for simultaneous fitting is immediately apparent if we look at the model prediction equations for MAOS material functions, Eqs. (19)-(22) . $G_i$ and $\alpha_i$ appear as a product ($G_i\,\alpha_i$) in the front factor, suggesting a strong correlation between the two and hence, larger uncertainties in both. An easy way to understand larger uncertainties due to correlation here is to see that a large increase in one parameter can be countered by a large decrease in the other parameter such that the product remains the same. Hence, each parameter can have a large variation for a given confidence level in model predictions. In the case of sequential fitting, $G_i$'s are fit from linear data and then fixed at those estimates in the equations of MAOS material functions. Hence, the strong correlations of $\alpha_i$'s are removed which then gives smaller apparent uncertainties (and similarly smaller apparent uncertainties are observed for $G_i$'s.) Hence, as discussed previously in Section 2.1, the model parameter uncertainties of sequential fitting reflect only a partial uncertainty which is smaller than the correct full uncertainty that accounts for all parameter correlations.

Even though $G_i$ and $α_i$ have larger uncertainties individually because of their inverse correlation, the product $G_i α_i$ has reduced uncertainty (i.e. determined with more confidence for simultaneous fitting as shown in Table 1). In fact, for the product $G_i α_i$, the relative uncertainties are smaller for simultaneous fitting than the sequential fitting, for two out of the three modes. This shows that while larger relative uncertainties are obtained in model parameters for simultaneous fitting compared to sequential fitting, other quantitates derived from model parameters can have smaller (or comparable) uncertainties as we further show next.

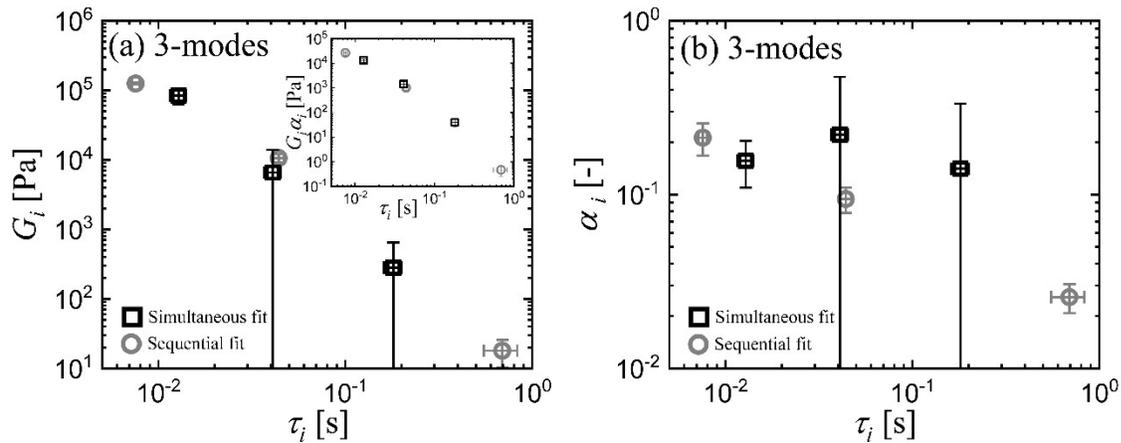

**FIG. 6.** The model parameter estimates and their uncertainties for a three-mode Giesekus model fit to the data in Fig. 5: (a) $G_i$ vs $τ_i$, and (b) $α_i$ vs $τ_i$. The inset in (a) shows the product $G_iα_i$ vs $τ_i$.

While Fig. 6 shows the comparison in fit parameters and their uncertainties for a multi-mode model, often we are interested in low-dimensional or average viscoelastic constants based on these parameters, which can then be related to microstructural properties. Four linear viscoelastic constants for a multi-mode Maxwell model are shown in the Appendix, Eqs. (27)-(30) : $G_0$, $η_0$, $\bar{τ}_1$, $\bar{τ}_2$ (The "bar" notation is used to distinguish these averages from the individual mode relaxation times). In addition, we derive a method to calculate average (or characteristic) nonlinear viscoelastic constants for a multi-mode Giesekus model as dimensionless parameters in the high frequency limit, Eqs. (32), (35) : $\bar{\alpha}_G$, $\bar{\alpha}_{G/\lambda}$. While the linear viscoelastic constants depend solely on the linear parameters, the nonlinear viscoelastic constants depend upon both linear and nonlinear model parameters.

Table 1. Fit parameters and their associated uncertainties for a three-mode Giesekus model using sequential and simultaneous fitting. Also shown are the estimates and uncertainties in quantities derived from model parameters.

| Parameter | Estimate ± Uncertainty | | Relative Uncertainty (%) | | |
|---|---|---|---|---|---|
| | Sequential | Simultaneous | Sequential | Simultaneous | Simultaneous / Sequential |
| $G_1$ [kPa] | 125±3.27 | 84.1±21.8 | 2.62 | 25.9 | 9.89 |
| $G_2$ [kPa] | 10.6±1.39 | 6.54±7.36 | 13.1 | 112.5 | 8.59 |
| $G_3$ [kPa] | 0.017±0.008 | 0.28±0.37 | 47.0 | 132.1 | 2.81 |
| $\tau_1$ [ms] | 7.56±0.36 | 12.8±1.25 | 4.76 | 9.76 | 2.05 |
| $\tau_2$ [ms] | 44.0±2.87 | 40.9±4.08 | 6.50 | 9.98 | 1.53 |
| $\tau_3$ [ms] | 692±140 | 181±19.5 | 20.2 | 10.7 | 0.52 |
| $\alpha_1$ [-] | 0.212±0.045 | 0.156±0.047 | 21.2 | 30.1 | 1.42 |
| $\alpha_2$ [-] | 0.094±0.016 | 0.221±0.255 | 17.02 | 115.4 | 6.78 |
| $\alpha_3$ [-] | 0.026±0.005 | 0.141±0.192 | 19.2 | 136.1 | 7.09 |
| $G_1\alpha_1$ [kPa] | 265±5.67 | 13.2±2.16 | 21.4 | 16.4 | 0.77 |
| $G_2\alpha_2$ [kPa] | 0.998±0.211 | 1.45±0.39 | 21.2 | 27.3 | 1.29 |

| $G_3\alpha_3$ [kPa] | $(0.460\pm0.217)$ $\times 10^{-3}$ | $0.039\pm0.014$ | 47.2 | 36.2 | 0.77 |
|---|---|---|---|---|---|
| $G_0$ [kPa] | $135\pm3.90$ | $91.0\pm17.1$ | 2.89 | 18.8 | 6.50 |
| $\eta_0$ [Pa-s] | $1423\pm8.072$ | $1398\pm131.7$ | 0.57 | 9.4 | 16.53 |
| $\bar{\tau}_1$ [ms] | $10.5\pm0.28$ | $15.4\pm2.21$ | 2.67 | 14.4 | 5.37 |
| $\bar{\tau}_2$ [ms] | $25.4\pm0.13$ | $24.3\pm0.80$ | 0.51 | 3.92 | 7.69 |
| $\bar{\alpha}_G$ [-] | $0.203\pm0.042$ | $0.161\pm0.035$ | 20.7 | 21.7 | 1.05 |
| $\bar{\alpha}_{G/\lambda}$ [-] | $0.210\pm0.044$ | $0.158\pm0.042$ | 21.0 | 26.6 | 1.27 |

A comparison between the estimates of viscoelastic constants and their uncertainties is shown in Table 1 and Fig. 7. For a number of constants, the two schemes yield significantly different results, i.e. non-overlapping values within one-standard deviation limit, specifically for $G_0$, $\bar{\tau}_1$, and $\bar{\tau}_2$. For these cases, the overall relative uncertainties are larger for simultaneous fit as one would expect. For the other three constants, the values overlap within their one-standard deviation limit. And for the nonlinear viscoelastic constants, in fact, the simultaneous fit relative uncertainties are comparable to sequential fit relative uncertainties as shown in Table 1. This is because the nonlinear viscoelastic constants involve the product $G_i\alpha_i$ which is determined with more confidence as discussed previously. These results show that the definition of derived quantities such as the low-dimensional viscoelastic constants can also have a huge impact on their estimates and uncertainties, along with the decision of choosing the fit scheme. However, it is always best to choose the simultaneous fitting to correctly account for all model parameter correlations, so that the uncertainties, whether small or large, are honestly acknowledged.

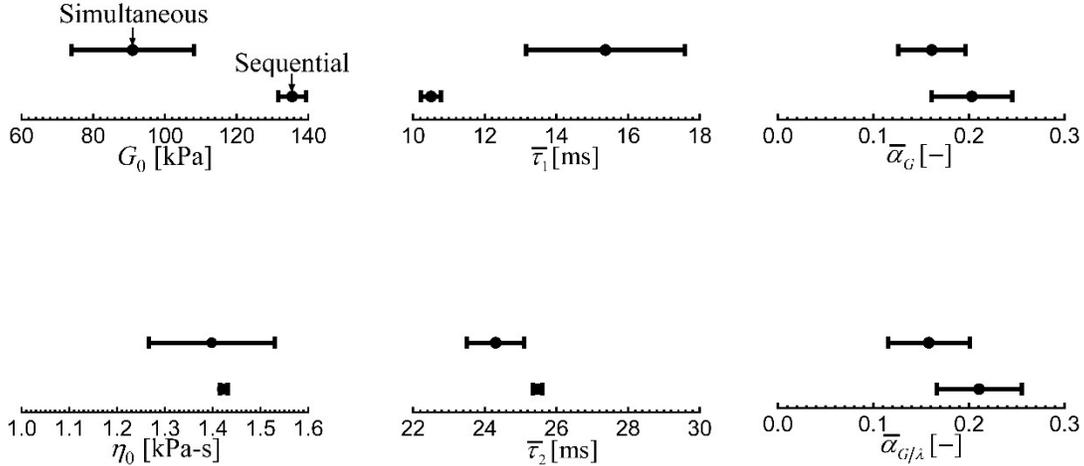

**FIG.7.** Estimates and uncertainties of the low-dimensional viscoelastic constants for the three-mode Giesekus model, as defined in Eqs.(27)-(30),(32), and (35).

## 5. Conclusions

We looked at the case of fitting linear and (weakly) nonlinear rheology data: SAOS and MAOS respectively, elucidating the effects of ignoring model parameter correlations. While fitting linear and nonlinear data sequentially (a common approach in rheology), important model parameter correlations are neglected. This gives a smaller relative uncertainty (or a false sense of certainty) in model parameters, which nonetheless is wrong and should be avoided whenever possible. Correct uncertainty in model parameters is obtained when both linear and nonlinear data are fit simultaneously, as shown for a three-mode Giesekus model fit in Fig. 8. The distinction between sequential and simultaneous fitting can be significant for model parameter uncertainties even when working with a weakly nonlinear dataset, where one might expect the linear data to dictate the fitting. Instead, we see that weakly nonlinear data provides significant information to inform both the nonlinear and the linear viscoelastic parameters of the relaxation spectrum.

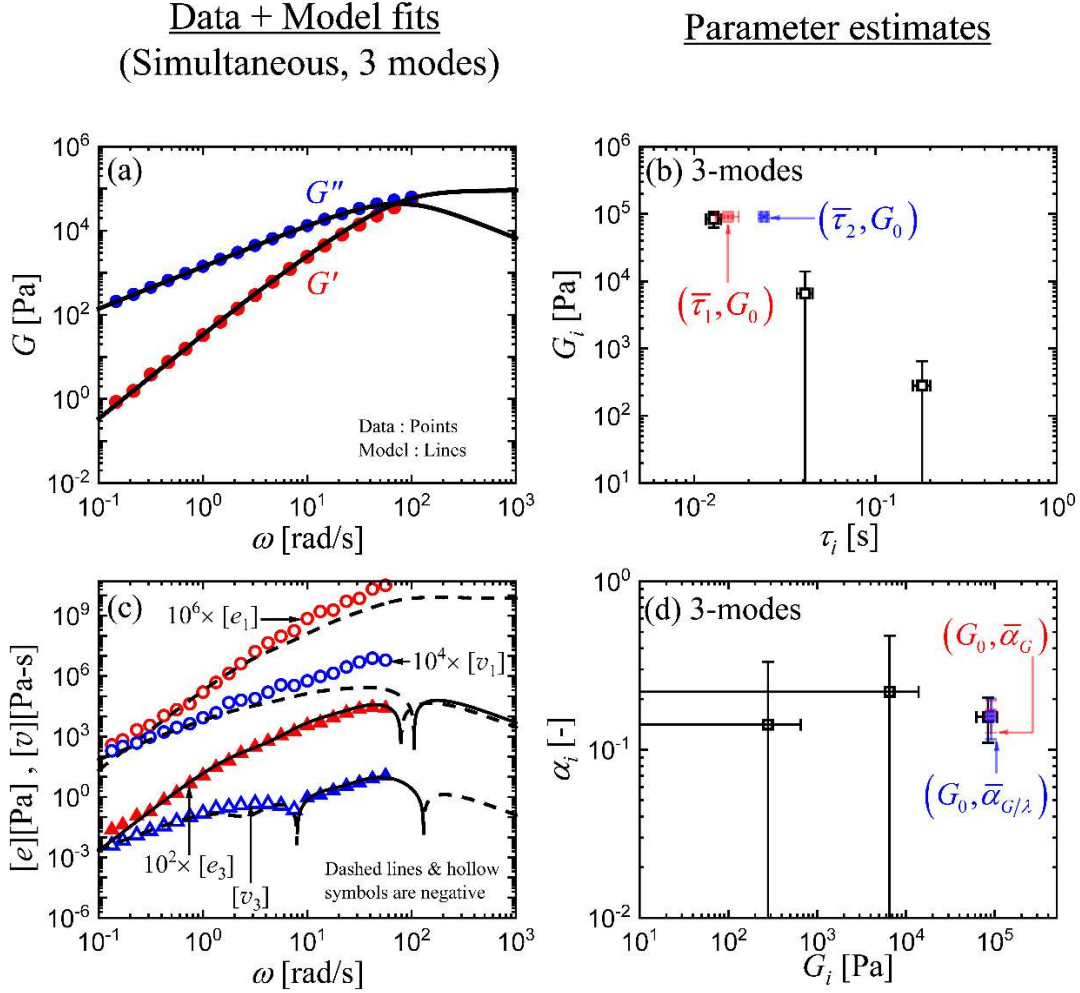

**FIG. 8.** A three-mode Giesekus model fit simultaneously to the linear (a) and nonlinear (c) data, accounting for all model parameter correlations. The estimates and associated uncertainties of model parameters, and a few low-dimensional viscoelastic constants (shown/derived in Appendix) are shown in (b) and (d).

While the relative uncertainties in model parameters can be significantly larger for simultaneous fitting because of the freedom allowed by the correlations, certain products of fit parameters might show minimal difference in their uncertainties from different fit methods. For the particular example shown here, this occurred for the nonlinear viscoelastic constants which were defined based on the product of a linear and a nonlinear model parameter. Here, while the individual parameters had increased uncertainties due to the inverse correlation, the product was determined with more confidence, and the distinction between sequential and simultaneous fitting was not that significant. However, such a result may be fortuitous and not suggestive of using sequential fitting in general.

Our results indicate the importance of simultaneous fitting of linear and nonlinear data for proper uncertainty quantification. Uncertainty will propagate through calculations, down to molecular scales to infer microstructural features, and up to macroscopic scales with predictive flow simulation [120]. Whenever possible, datasets should be fit simultaneously to estimate the correct model parameter uncertainty accounting for all correlations. When simultaneous fitting cannot be used, then the unknown missing parameter uncertainty should be acknowledged.

**Supplementary Material**

The supplementary material shows model fits comparison between sequential and simultaneous fitting for number of modes, $K = 1, 2, 4, 5$.

**Acknowledgements**

We acknowledge the generous support of Dr. N. Ashwin Bharadwaj for providing the experimental dataset used in this work. This work was financially supported by the ACS Petroleum Research Fund and by the U.S. Department of Energy, Office of Basic Energy Sciences, Division of Materials Sciences and Engineering under Award No. DE-SC0020858, through the Materials Research Laboratory at the University of Illinois Urbana-Champaign.

**Appendix: Derivation of characteristic (average) α for a multi-mode Giesekus model**

**A.1. Linear viscoelastic constants**

A low-dimensional modulus and viscosity can be defined based on the relaxation spectrum as [121,122]:

$$G_0 = \lim_{De(Shortest) \to \infty} G'(\omega) = \sum_{i=1}^{K} G_i \tag{27}$$

$$\eta_0 = \lim_{De(Longest) \to 0} \eta'(\omega) = \sum_{i=1}^{K} G_i \lambda_i. \tag{28}$$

Here De(Shortest) and De(Longest) refer to Deborah numbers based on the shortest and longest relaxation modes in the spectrum, respectively. In addition, two time-scales are defined as follows [121,122]:

$$\bar{\tau}_1 = \frac{\lim_{De(Longest) \to 0} \eta'(\omega)}{\lim_{De(Shortest) \to \infty} G'(\omega)} = \frac{\sum_{i=1}^{K} G_i \lambda_i}{\sum_{i=1}^{K} G_i} = \frac{\eta_0}{G_0} \tag{29}$$

$$\bar{\tau}_2 = \frac{\sum_{i=1}^{K} G_i \lambda_i^2}{\sum_{i=1}^{K} G_i \lambda_i} \tag{30}$$

The original definitions in Eqs. (29)-(30) do not involve "bar" notation. We have introduced the bars to distinguish the two time scales from the individual relaxation times of the various modes of a multi-mode model.

**A.2 Nonlinear viscoelastic constants**

While the use of linear viscoelastic constants in Eqs. (27)-(30) is very common, there is no set protocol to define low-dimensional nonlinear viscoelastic constants. Here, we derive two new nonlinear viscoelastic constants for the multi-mode Giesekus model based on moments of the spectra related to the high-frequency MAOS limit.

The asymptotic limits of MOAS material functions for a Giesekus model, as derived by Bharadwaj and Ewoldt [30], provide a means for defining appropriate low-dimensional nonlinear viscoelastic constants similar to linear viscoelastic constants of Eqs. (27)-(30). Table 2 and 3 of Bharadwaj and Ewoldt show the expression of elastic and viscous MAOS material functions in the high and low frequency limits. While some limits have quadratic terms in $\alpha$, we focus on the limits with linear dependence on $\alpha$. One appropriate limit is for $[e_1]$ at large frequencies, which after generalizing to a multi-mode description gives

$$-2 \times \lim_{\text{De(Shortest)} \to \infty} [e_1](\omega) = \sum_{i=1}^{K} \alpha_i G_i . \tag{31}$$

Eq. (31) can be normalized by (27) to give a constant in the high frequency limit, which is essentially the average of $\alpha$'s weighted by $G$'s and meaningful in the high-frequency elastic MAOS response. We call this $\bar{\alpha}_G$,

$$\bar{\alpha}_G = -2 \times \frac{\lim_{\text{De(Shortest)} \to \infty} [e_1](\omega)}{\lim_{\text{De(Shortest)} \to \infty} G'(\omega)} = \frac{\sum_{i=1}^{K} \alpha_i G_i}{\sum_{i=1}^{K} G_i} . \tag{32}$$

Another useful limit is for $[v_3]$ at large frequencies, which after generalizing to a multi-mode description gives

$$-8 \times \lim_{\text{De(Shortest)} \to \infty} [v_3](\omega) = \sum_{i=1}^{K} \frac{\alpha_i G_i}{\lambda_i \omega^2} . \tag{33}$$

To normalize Eq. (33), we use the limit of $\eta'$ at large frequencies

$$\lim_{\mathrm{De(Shortest)}\to\infty} \eta'(\omega) = \sum_{i=1}^{K} \frac{G_i}{\lambda_i \omega^2}. \tag{34}$$

Eq. (33) can be normalized by (34) to give another constant in the high frequency limit, which is essentially the average of $\alpha$'s weighted by $G/\lambda$'s and meaningful in the high-frequency viscous MAOS response. We call this $\bar{\alpha}_{G/\lambda}$,

$$\bar{\alpha}_{G/\lambda} = -8 \times \frac{\lim_{\mathrm{De(Shortest)}\to\infty} [v_3](\omega)}{\lim_{\mathrm{De(Shortest)}\to\infty} \eta'(\omega)} = \frac{\sum_{i=1}^{K} \alpha_i G_i / \lambda_i}{\sum_{i=1}^{K} G_i / \lambda_i}. \tag{35}$$

Eqs. (32) and (35) provide the low-dimensional nonlinear viscoelastic constants for a multi-mode Giesekus model. Interestingly, they are calculated in the limit of high frequencies. In contrast, the linear viscoelastic constants in Eqs. (27)-(30) are calculated for both the low and high frequency limits.

# Supplementary Material

# On simultaneous fitting of nonlinear and linear rheology data: Preventing a false sense of certainty

Piyush K. Singh [*], Randy H. Ewoldt [†]

Department of Mechanical Science and Engineering, University of Illinois Urbana-Champaign, Urbana, IL 61801, USA

**SM1. Model fit comparison between sequential and simultaneous fitting**

Shown below are the model fit comparison between sequential and simultaneous fitting for number of modes $K$ = 1, 2, 4, 5. The fit comparison for three-modes is shown in Fig. 5 of the main manuscript.

[*] Current affiliation: Department of Chemical and Biomolecular Engineering, University of Illinois at Urbana-Champaign, Urbana, IL 61801, USA

[†] Corresponding author: ewoldt@illinois.edu

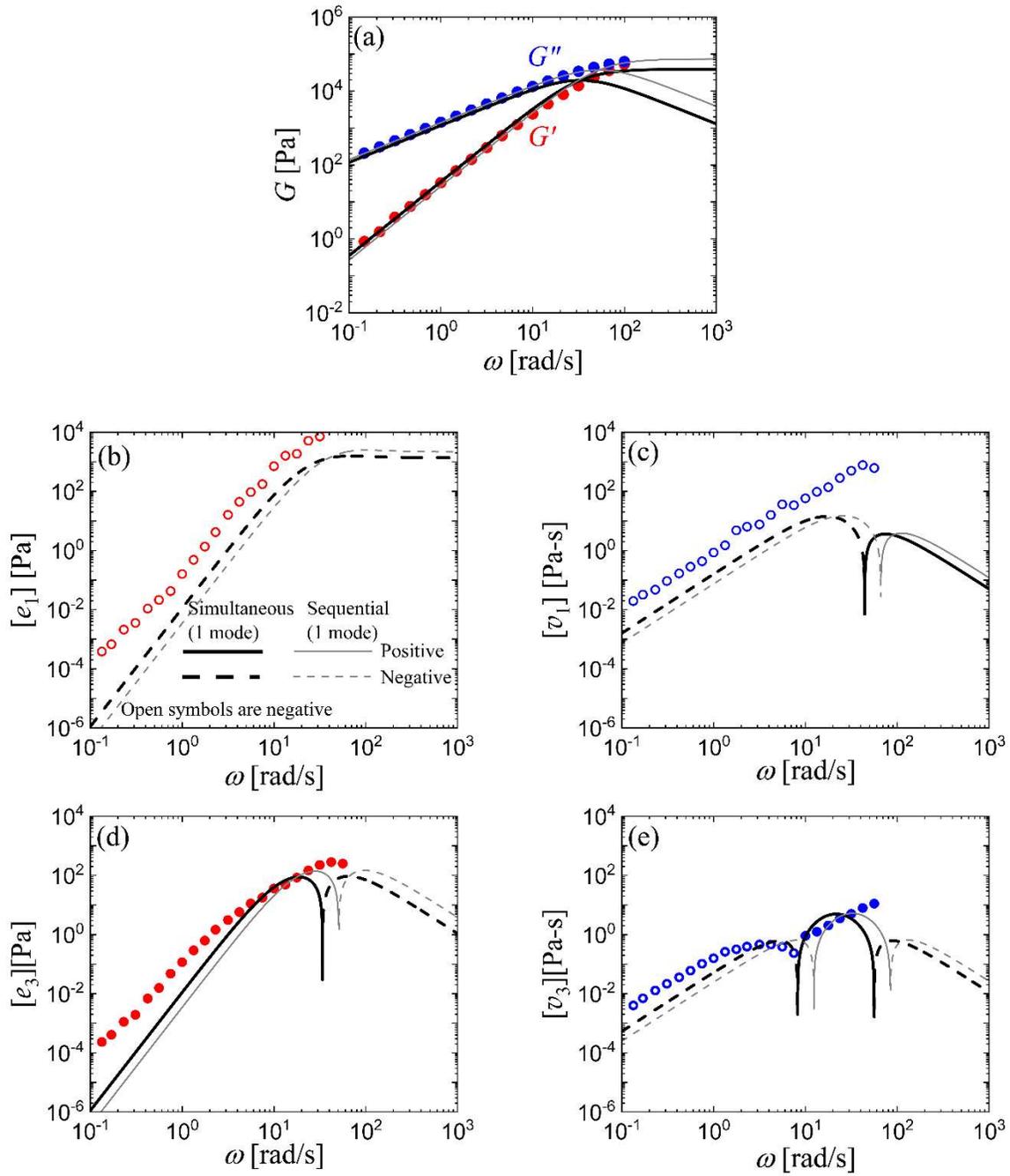

**FIG. SM1.** Single-mode Giesekus model fit to the data of Fig. 3 of the main manuscript for both sequential and simultaneous fit schemes.

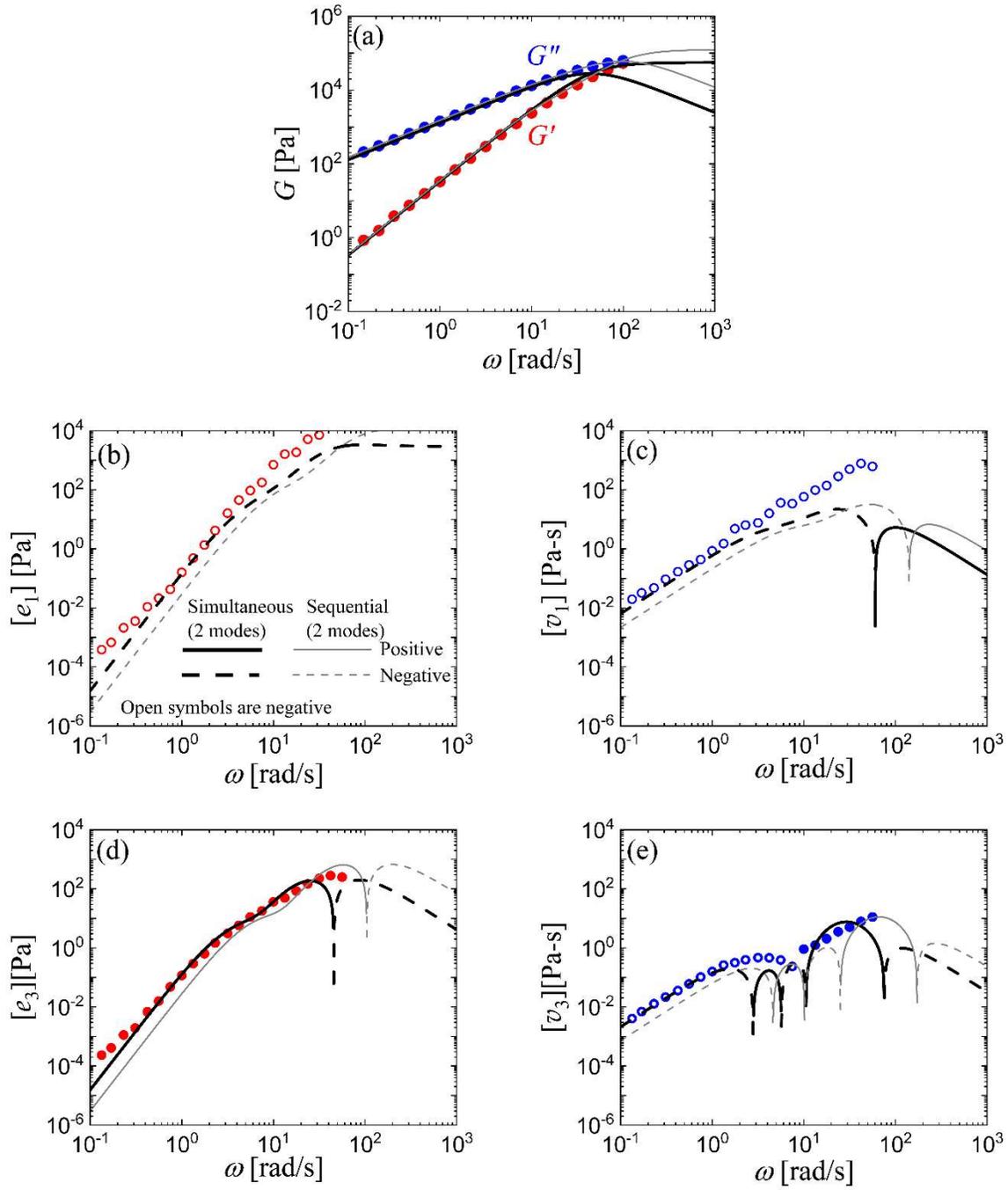

**FIG. SM2.** Two-mode Giesekus model fit to the data of Fig. 3 of the main manuscript for both sequential and simultaneous fit schemes.

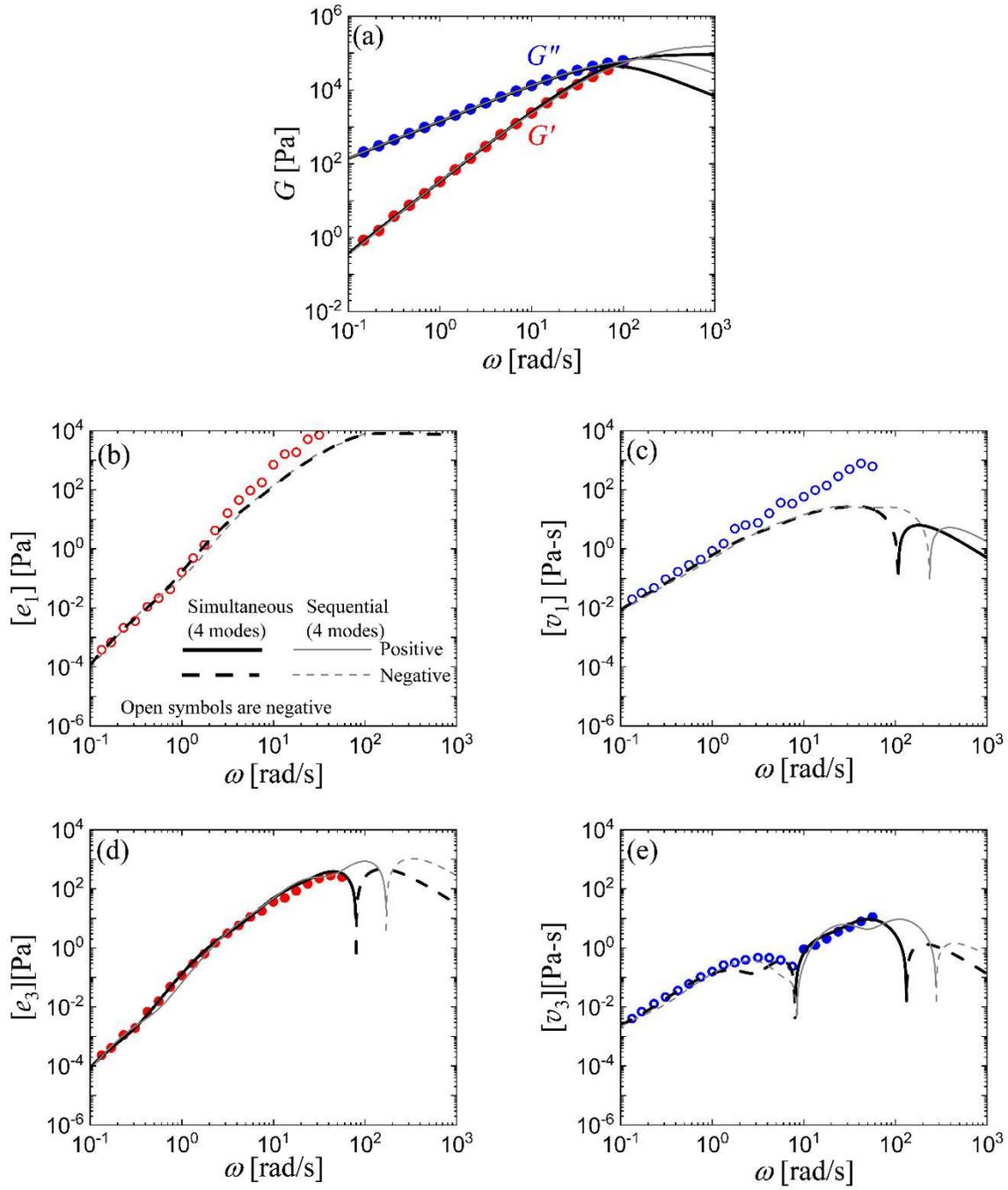

**FIG. SM3.** Four-mode Giesekus model fit to the data of Fig. 3 of the main manuscript for both sequential and simultaneous fit schemes.

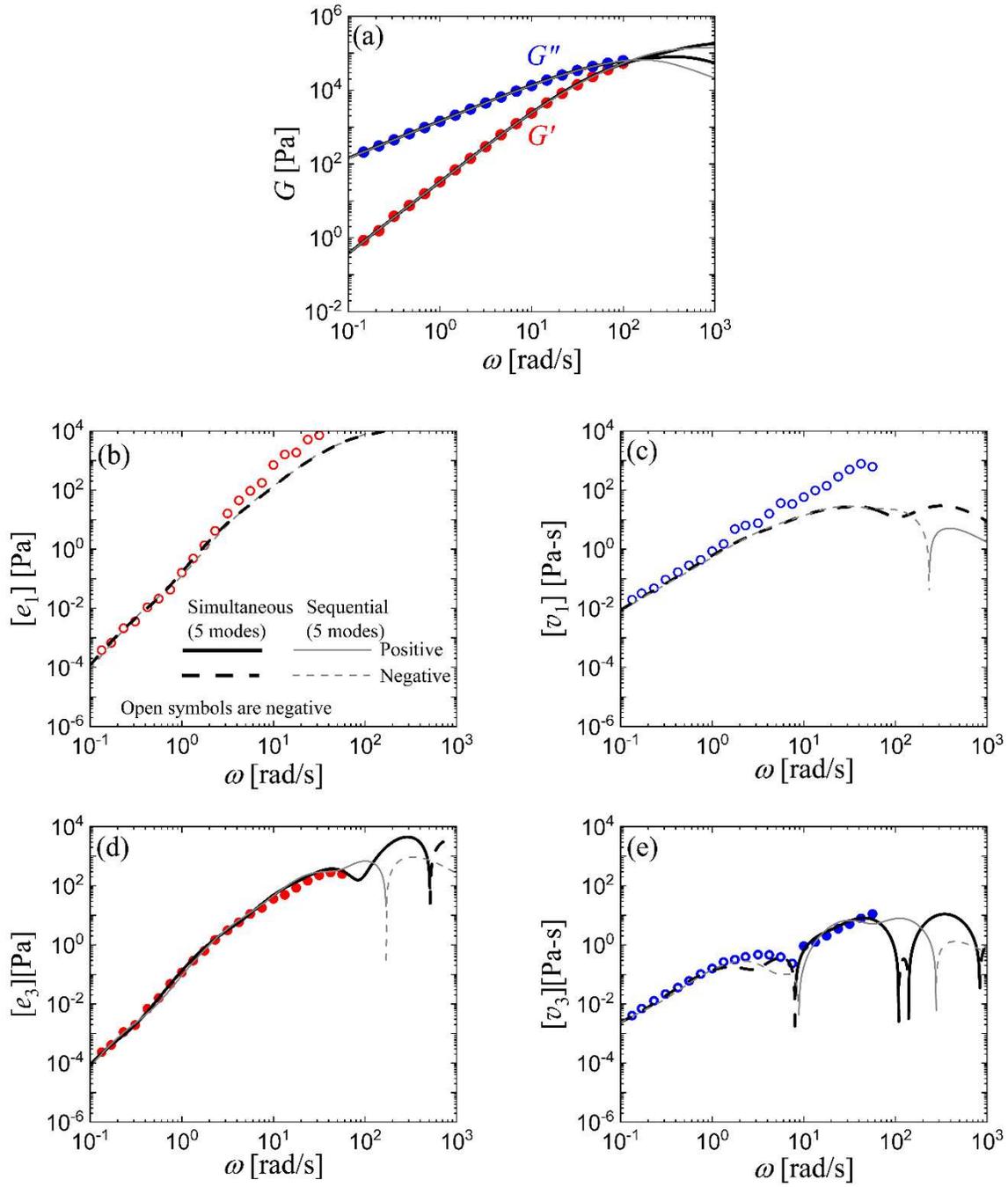

**FIG. SM4.** Five-mode Giesekus model fit to the data of Fig. 3 of the main manuscript for both sequential and simultaneous fit schemes.